\documentclass[aps,prb,twocolumn,superscriptaddress, show pacs]{revtex4-1}
\usepackage{bm, amsmath}
\usepackage{graphicx}
\usepackage{color}
\usepackage{epstopdf}
\newcommand{\blue}[1]{\textcolor{blue}{#1}} 
\usepackage{graphicx}
\usepackage{dcolumn}
\usepackage{bm}
\usepackage{subfigure}

\begin{document}

\title{Chiral surface twists and skyrmion stability in nanolayers of cubic helimagnets}

\author{A.~O.~Leonov}
\thanks{ A.Leonov@ifw-dresden.de}
\affiliation{Center for Chiral Science, Hiroshima University, Higashi-Hiroshima, 
Hiroshima 739-8526, Japan}
\affiliation{IFW Dresden, Postfach 270016, D-01171 Dresden, Germany}   

\author{Y. Togawa} 
\affiliation{Center for Chiral Science, Hiroshima University, Higashi-Hiroshima, 
Hiroshima 739-8526, Japan}
\affiliation{Osaka Prefecture University, 1-2 Gakuencho, Sakai, Osaka 599-8570, Japan}   

\author{T.~L.~Monchesky}
\affiliation{Department of Physics and Atmospheric Science, Dalhousie University, 
Halifax, Nova Scotia, Canada B3H 3J5}

\author{A.~N.~Bogdanov}
\affiliation{Center for Chiral Science, Hiroshima University, Higashi-Hiroshima, 
Hiroshima 739-8526, Japan}
\affiliation{IFW Dresden, Postfach 270016, D-01171 Dresden, Germany} 

\author{J.~Kishine}
 \affiliation{Center for Chiral Science, Hiroshima University, Higashi-Hiroshima, 
Hiroshima 739-8526, Japan}
\affiliation{The Open University of Japan, Chiba 261-8586, Japan}

\author{Y.~Kousaka}
 \affiliation{Center for Chiral Science, Hiroshima University, 
Higashi-Hiroshima, Hiroshima 739-8526, Japan}

\author{M.~Miyagawa}
 \affiliation{Center for Chiral Science, Hiroshima University, 
Higashi-Hiroshima, Hiroshima 739-8526, Japan}

\author{T.~Koyama}
 \affiliation{Center for Chiral Science, Hiroshima University, 
Higashi-Hiroshima, Hiroshima 739-8526, Japan}

\author{J.~Akimitsu}
 \affiliation{Center for Chiral Science, Hiroshima University, 
Higashi-Hiroshima, Hiroshima 739-8526, Japan}

\author{Ts. Koyama} 
\affiliation{Osaka Prefecture University, 1-2 Gakuencho, Sakai, Osaka 599-8570, Japan} 

  \author{K. Harada} 
\affiliation{Osaka Prefecture University, 1-2 Gakuencho, Sakai, Osaka 599-8570, Japan}   

\author{S. Mori} 
\affiliation{Osaka Prefecture University, 1-2 Gakuencho, Sakai, Osaka 599-8570, Japan}

\author{D. McGrouther} 
\affiliation{School  of  Physics  and  Astronomy, University  of  Glasgow,  Glasgow,  UK,  G12  8QQ}   

 \author{R. Lamb} 
\affiliation{School  of  Physics  and  Astronomy, University  of  Glasgow,  Glasgow,  UK,  G12  8QQ} 

  \author{M. Krajnak} 
\affiliation{School  of  Physics  and  Astronomy, University  of  Glasgow,  Glasgow,  UK,  G12  8QQ} 

 \author{S. McVitie} 
\affiliation{School  of  Physics  and  Astronomy, University  of  Glasgow,  Glasgow,  UK,  G12  8QQ}  

\author{R. L. Stamps} 
\affiliation{School  of  Physics  and  Astronomy, University  of  Glasgow,  Glasgow,  UK,  G12  8QQ}  

\author{K.~Inoue}
 \affiliation{Center for Chiral Science, Hiroshima University, 
Higashi-Hiroshima, Hiroshima 739-8526, Japan}

\date{\today}

\begin{abstract}
{
 Lorentz transmission electron microscopy (LTEM) 
investigations of modulated states in a FeGe wedge and detailed calculations 
demonstrate that chiral twists arising near the surfaces of  noncentrosymmetric 
ferromagnets (Meynell et al. Phys. Rev. B, \textbf{90}, 014406 (2014))
provide a stabilization mechanism for skyrmion lattices and helicoids
in cubic helimagnet nanolayers.
The calculated magnetic phase diagram for free standing cubic helimagnet 
nanolayers shows that magnetization processes in these compounds
fundamentally differ from those in bulk cubic helimagnets and are
characterized by the first-order transitions between 
modulated phases and the formation of specific multidomain states.
The paper reports LTEM observations of multidomain patterns
in FeGe free-standing nanolayers. 
}
\end{abstract}

\pacs{
75.30.Kz, 
12.39.Dc, 
75.70.-i.
}
         
\maketitle

\vspace{5mm}

\section{Introduction}

 \textit{Dzyaloshiskii-Moriya} (DM) interactions  \cite{Dz64} 
stabilize two-dimensional axisymmetric solitonic states 
(\textit{chiral skyrmions}) in saturated phases of  
magnetic materials with broken inversion symmetry 
\cite{JETP89,JMMM94}. 
In uniaxial noncentrosymmetric ferromagnets 
chiral skyrmions condense into hexagonal lattices below a
certain critical field and remain thermodynamically stable 
(correspond to the global minimum of the magnetic energy 
functional) in a broad range of applied magnetic fields  
\cite{JMMM94}.
This does not occur in bulk cubic helimagnets where
one-dimensional modulations along the applied field 
(the \textit{cone} phase) \cite{Bak80} have the lowest energy
 practically in the whole area of the magnetic phase diagram, 
and skyrmion lattices can exist only as metastable states 
\cite{Butenko10,Wilson14}.

During last years numerous observations 
of different types of skyrmion states have been reported
 in free-standing nanolayers and epilayers 
of cubic helimagnets 
(e.g. \cite{Yu10,Yu11,Yu15,Huang12,Wilson12,Yokouchi15}).
These findings have given rise to a puzzling question:
why are skyrmion lattices totally suppressed in bulk
cubic helimagnets  but easily arise in nanolayers of  the same
compounds?

Two physical mechanisms have been proposed to date to explain
the formation of skyrmion lattices in confined cubic helimagnets.
One of them is based on effects imposed   by induced uniaxial 
anisotropy \cite{Butenko10,Wilson14}.   
In epilayers of cubic helimagnets on Si (111) substrates,
a strong uniaxial anistropy is induced by the lattice mismatch
between the B20 crystal and the substrate \cite{Karhu10,Wilson14}.
This uniaxial anisotropy suppresses the cone phase
and stabilizes a number of nontrivial chiral modulated states
including out-of-plane and in-plane skyrmion lattices recently
observed in cubic helimagnet epilayers
\cite{Wilson12,Huang12,Wilson14}. 

The second stabilization mechanism  is provided by 
specific modulations (\textit{chiral twists}) arising
near the surfaces of confined cubic helimagnets  
 \cite{Rybakov13,Wilson13,Meynell14}.
Chiral twists have been recently discovered in MnSi/Si(111) 
films \cite{Wilson13,Meynell14}. However, their influence on
the magnetic states arising in confined cubic helimagnets
is still unclear. 
Also, physical mechanisms underlying
the formation of skyrmionic states in free standing
films of cubic helimagnets are  unknown and 
a theoretical description of 
arising magnetic states in these systems
 is still an open question.

In this paper we report LTEM investigations of
modulated states in a FeGe wedge and theoretical analysis
of magnetic states in confined cubic helimagnets.
Our findings show that surface twist 
instabilities play a decisive role in the stabilization
of skyrmionic states in free standing layers of cubic 
helimagnets.



\section{Phenomenological model and magnetic phases}

\subsection{Model}

The standard model for magnetic states
in cubic non-centrosymmetric ferromagnets is based on
the energy density functional \cite{Dz64,Bak80}
\begin{equation}
w =A\,(\mathbf{grad}\,\mathbf{m})^2 + 
D\,\mathbf{m}\cdot \mathrm{rot}\,\mathbf{m}
-\mu_0 \,M  \mathbf{m} \cdot \mathbf{H},
\label{density}
\end{equation}
including the principal interactions essential 
to stabilize modulated states: 
the exchange stiffness with constant $A$, 
Dzyaloshinskii-Moriya (DM) coupling energy
with constant $D$, and the Zeeman energy;
\begin{eqnarray}
\mathbf{m}= (\sin\theta\cos\psi;\sin\theta\sin\psi;\cos\theta) 
\label{m}
\end{eqnarray}
is the unity vector along the magnetization vector
 $\mathbf{M} = \mathbf{m} M$, and $\mathbf{H}$ is
the applied magnetic field.

\begin{figure}
\includegraphics[width=1.0\columnwidth]{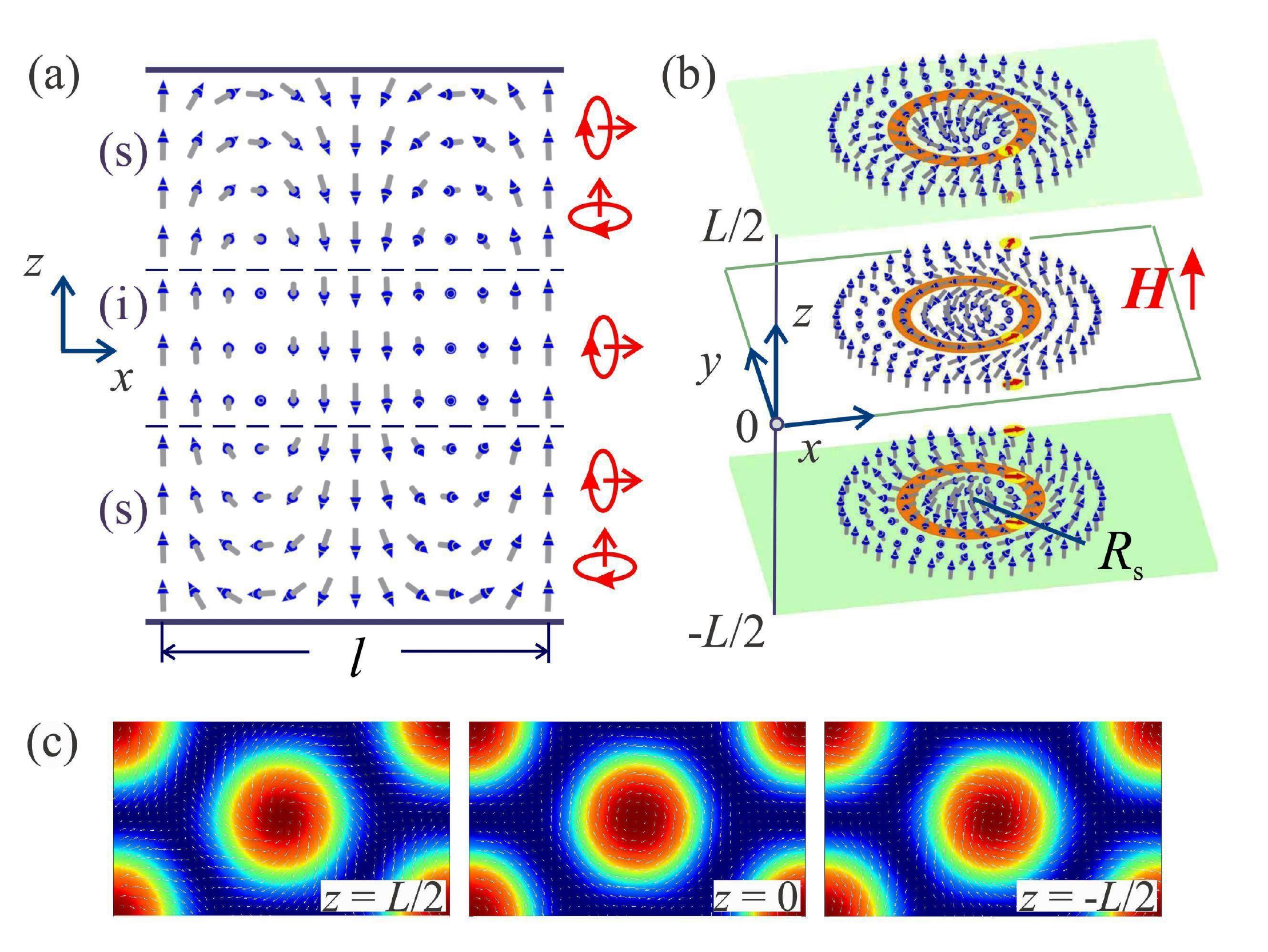}
\caption{ 
(color online). Magnetic structure  of a helicoid 
with   period $l$ (a) and a skyrmion lattice cell of
radius $R_s$ (b,c) in nanolayers of cubic helimagnets.
In the internal area (i) the helicoid has 
in-plane modulations along the $x$-axis, the surface areas (s)
are modulated along the $x$  and $z$ axes.
\label{fig:pics}
}
\end{figure}

We investigate  the functional (\ref{density}) 
in a film of thickness $L$ infinite in $x-$ 
and $y-$ directions and confined by parallel planes at $z = \pm L/2$ 
in magnetic field $\mathbf{H}$ applied along $z-$ axis 
(Fig. \ref{fig:pics} a). 

The equilibrium magnetic states in the film are derived 
by the Euler equations for energy functional (\ref{density}) 
together with the Maxwell equations and with corresponding
boundary conditions.
The solutions depend on the two control parameters 
of the model (\ref{density}), the \textit{confinement ratio}, 
$\nu $ and the reduced value of the applied magnetic field, $h $
\begin{eqnarray}
\nu = \frac{L}{L_D}, \ h =\frac{H}{H_D}, \
 L_D = \frac{4\pi A}{|D|}, \ \mu_0 H_D = \frac{D^2}{2A M} \ 
\label{period}
\end{eqnarray}
where $L_D$ is the \textit{helix period} 
and $H_D$ is the \textit{saturation field} 
\cite{Bak80,JMMM94}. 

\subsection{ Modulated states in bulk cubic helimagnets}
Magnetic states in bulk cubic helimagnets are commonly described
by \textit{unconfined} modulated states including
the following three phases \cite{Dz64,Bak80,JMMM94}:

\textbf{(i)} \textit{Cones} are chiral single-harmonic modulations 
along the applied field. The solutions for the cone phase and
the equilibrium energy density are derived in analytical form \cite{Bak80}
\begin{eqnarray}
\cos \theta_c =  h, \quad \psi_c = 2\pi z/L_D, \quad
 w_c (h) = -K_0 \left( 1+h^2 \right) \
\label{cone}
\end{eqnarray}
where $K_0 = D^2/(4 A) = \mu_0 H_D M/2$ is the effective easy-plane anisotropy
imposed by the cone modulations \cite{Butenko10,Wilson14}.

\textbf{(ii)} \textit{Helicoids} are one-dimensional chiral modulations 
with the propagation direction perpendicular to the applied field 
and homogeneous along the direction of the applied field  \cite{Dz64}.
Helicoids propagating along the $x$-axis are described by solutions 
($\theta (x), \psi = \pi/2$).
The Euler equation for the helicoid energy density 
\begin{eqnarray}
w_h^{0} (\theta) =  A \,\theta_x^2 - D \,\theta_x - \mu_0 M H \cos \theta
\label{helicoid0}
\end{eqnarray}
yields a set of parametrized periodic solutions $\theta (x,l)$
where the parameter $l$ designates the period of helicoids.
The equilibrium period $l_0$ and profile $\theta (x, l_0)$ are derived
by minimization of the helicoid energy density with respect 
to $l$  \cite{Dz64}.

\textbf{(iii)} \textit{Skyrmion lattices}. 
The axisymmetric cores of chiral skyrmion 
lattice cells are described by solutions
\cite{JMMM94}
\begin{eqnarray}
\theta (\rho), \quad \psi = \pi/2 + \varphi
\label{skyrmioncell0}
\end{eqnarray}
where $\mathbf{r} = (\rho \cos \varphi, \rho \sin \varphi, z)$ 
are cylindrical coordinates of the spatial variable.

The equilibrium periods and magnetization profiles $\theta (\rho)$ of the
skyrmion lattice cells are derived by minimization of the energy density 
functional \cite{JMMM94} 
 \begin{eqnarray}
\textstyle
 w_s^{0} (\theta)   =   A \: \mathcal{J}_s^{0} ( \theta )   
+ D \:\mathcal{I}_s^{0} ( \theta ) -  \mu_0 M H \cos \theta, \quad  
\label{skyrmion0}
\end{eqnarray}
$\mathcal{J}_s^{0} ( \theta )   =  \theta_{\rho}^2  +\frac{1}{\rho^2}\sin^2 \theta, \:$
$\mathcal{I}_s^{0} ( \theta )  = \theta_{\rho}+ \frac{1}{\rho} \sin \theta \cos \theta$ 
for different values of the core radii $R$, and optimization of the mean
energy density of the skyrmion lattice with respect to $R$.

Among these solutions, the cone phase (\ref{cone}) corresponds to
the global minimum of model (\ref{density}) over the whole region
where chiral modulations occur ($H < H_D$). The helicoids 
and skyrmion lattices  exist  as metastable states below the critical fields
$H_h = 0.617 H_D$ and $H_s = 0.801 H_D$ correspondingly 
\cite{Dz64,JMMM94}. 

\begin{figure}
\includegraphics[width=1.0\columnwidth]{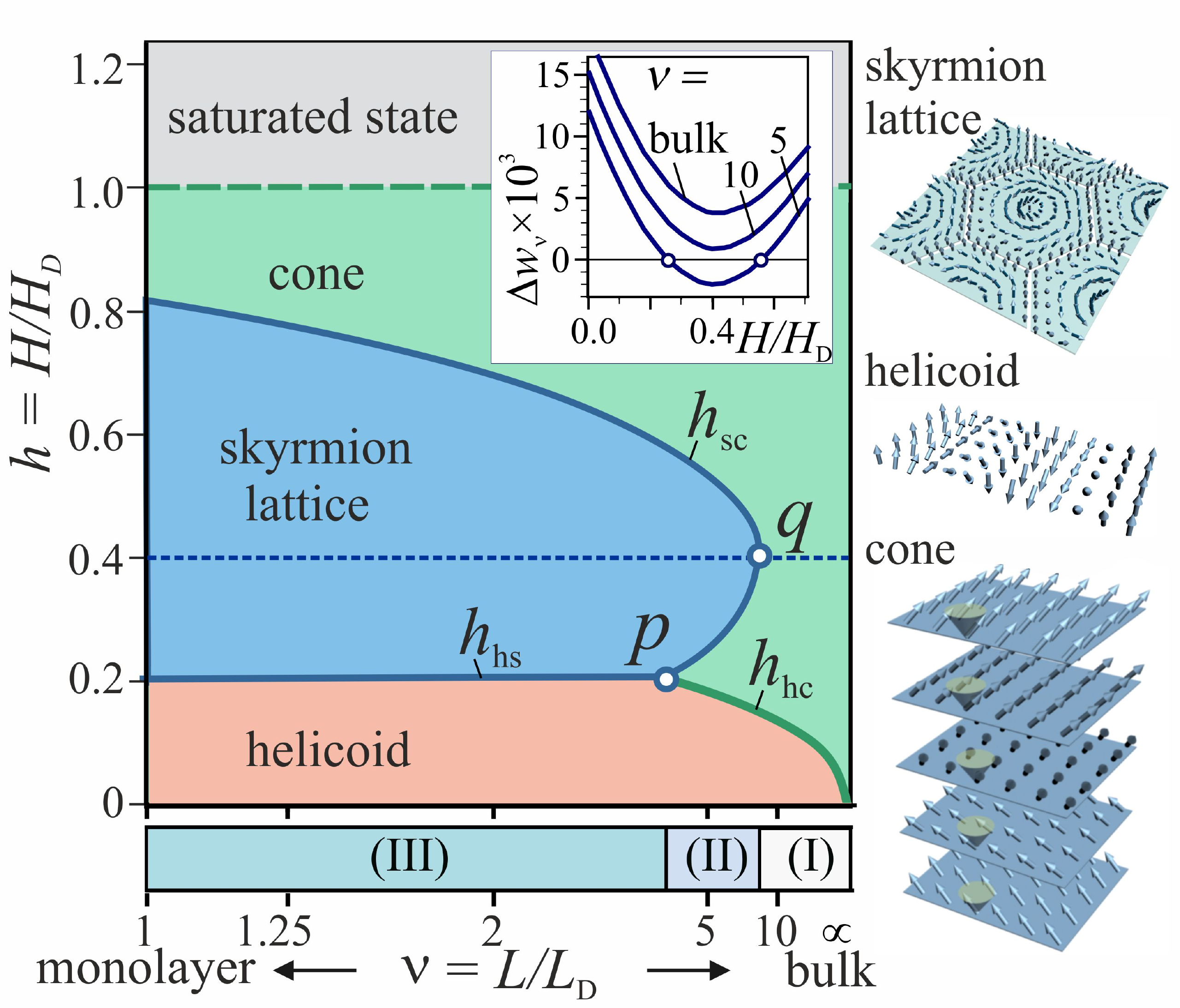}
\caption{ 
(color online) The magnetic phase diagram  of the magnetic states corresponding
to the global minima for model (1) in reduced variables for the film
thickness $\nu = L/L_D$ and applied magnetic field $h = H/H_D$.
The existence areas of the modulated phases (\textit{cone}, 
\textit{helicoids}, and \textit{skyrmion lattice}) are separated
by the first-order transition lines (solid). $p$ (4.47, 0.232) 
is a triple point, $q$ (7.56, 0.40) is a completion point.
Dashed line indicates the second-order transition between 
the cone and saturated state.
 Along the dotted line $H_a = 0.4 H_D$, the difference between the
energy densities of the skyrmion lattice and the cone phase 
($\Delta w_{\nu}$) is minimal (Inset).
\label{fig:phasediagram}
}
\end{figure}

\subsection{ Modulated states in confined cubic helimagnets}

The solutions for unconfined helicoids (\ref{helicoid0}) 
and skyrmion lattices (\ref{skyrmioncell0}) homogeneous along
the film normal describe magnetic configurations in the depth 
of a bulk cubic helimagnet.
However, the situation changes radically near the film surfaces. 
The gradient term, 
$$m_x\partial m_y/\partial z - m_y \partial m_x/\partial z$$
 in the DM energy functional  (Eq. (\ref{density})) violates transversal 
homogeneity of helicoids and skyrmion states and imposes chiral modulations 
along the $z-$ axis that decay into the depth of the sample (\textit{surface twists})
\cite{Wilson13,Rybakov13,Wilson14}. 
The \textit{penetration depth} of these surface modulations is comparable
with the characteristic length $L_D$ \cite{Wilson14}. 

Mathematically,  axisymmetric skyrmion cells in thin films 
 are described by solutions of type
$\theta = \theta (\rho, z), \psi = \psi (\varphi, z)$, and
helicoids propagating in a film along the $x$-axis are described by solutions of type
$\theta (x, z), \psi (x, z)$.

The energy density functional for confined helicoids ($w_h (\theta, \psi)$) 
and skyrmion lattices ($w_s (\theta, \psi)$) can be written  in the following 
form
\begin{eqnarray}
\!w_{h(s)} \! = \! A \mathcal{J}_{h(s)}\!( \theta, \psi) \!
+\! D \mathcal{I}_{h(s)}\!( \theta, \psi)\!-\!\mu_0 M\!H\!\cos\!\theta,\ 
\label{energy2}
\end{eqnarray}
where the  exchange ($\mathcal{J}_{h(s)}$) and Dzyaloshinskii-Moriya 
($\mathcal{I}_{h(s)}$) energy functionals read as

$ \mathcal{J}_{h} ( \theta, \psi)   =  \theta_x^2 + \theta_z^2 + 
\sin^2 \theta \left( \psi_x^2 + \psi_z^2 \right) $,

$\mathcal{I}_h ( \theta, \psi) =  \cos \psi \theta_x 
+ \sin \theta \cos \theta \sin \psi \ \psi_x
+ \sin^2 \theta \: \psi_z $,

$\mathcal{J}_s ( \theta, \psi)   =  \theta_{\rho}^2 + \theta_z^2 + 
\sin^2 \theta \left(\frac{1}{\rho^2}  \psi_{\varphi}^2 + \psi_z^2 \right) $,

$\mathcal{I}_s ( \theta, \psi) =  \sin (\psi- \varphi)
 ( \theta_{\rho}+ \frac{1}{\rho} \sin \theta \cos \theta \: \psi_{\varphi}) 
+ \sin^2 \theta \: \psi_z $.

The equilibrium solutions for confined helicoids and skyrmion lattices
are derived by solving the Euler equations for functional (\ref{energy2})  
with free boundary conditions at the film surfaces ($z = \pm L/2$).

Most of the investigated free standing films and epilayers 
of cubic helimagnets have a thickness 
exceeding the period of the helix ($L \geq L_D$).
In this paper we carry out  detailed analysis of the solutions for 
confined chiral modulations in cubic helimagnetic films with the thickness
ranging from $L = L_D$ to a bulk limit ($ L \gg L_D$).

\section{Magnetic phase diagram}

\subsection{Results of numerical simulations}

 The calculated $\nu-h$ phase diagram in Fig. \ref{fig:phasediagram}
indicates the areas with the chiral modulated states
corresponding to the global minimum of the energy functional and separated
by the first-order transition lines.
For $L \gg L_D$ the solutions for confined helicoids and
skyrmion lattices approach the solutions for the magnetic 
states in the unconfined case(\ref{helicoid0}), 
(\ref{skyrmioncell0}), which are homogeneous along the 
$z$-axis \cite{Dz64,JMMM94}.
Surface twist instabilities arising in confined cubic 
helimagnets \cite{Rybakov13,Meynell14}  provide a thermodynamical stability 
for helicoids and skyrmion lattices in a broad range of the applied fields 
(Fig. \ref{fig:phasediagram}). 

Another noticeable feature of the phase diagram is that the line
$h = 0.4$ is a symmetry axis for the skyrmion lattice stability area.
This follows from the fact that in bulk helimagnets this field corresponds 
to the minimal value of the skyrmion lattice energy compared to that of the cone phase
\cite{Butenko10,Wilson14}. This effect plays a crucial role in the formation of the A-phase
pocket near the ordering temperature of bulk cubic helimagnets (for details see the Ref.  \cite{Wilson14}).
The differences between the equilibrium average energy densities of the skyrmion lattice 
($\bar{w}_{\mathrm{s}}$) and the energy density of the cone phase ($w_{\mathrm{c}}$)
$\Delta w_{\nu} (h) = \bar{w}_{\mathrm{s}} (h, \nu)-w_{\mathrm{c}}(h, \nu)$ plotted
as functions of the applied field also reach the minimum in the fields close  to $h = 0.4$ 
(Inset of Fig. \ref{fig:phasediagram}). As a result, below $\nu_q = 7.56$, the stability 
area of the skyrmion lattices extends around the line $h = 0.4$.

In the whole range of the film thickness, the helicoids with in-plane propagation 
directions correspond to the group state of the system. 
The triple point $p$ (4.47, 0.232) and the completion point $q$ (7.56, 0.40)
split the phase diagram into three distinct areas with different types of 
the magnetization processes.

(I) $ \nu > \nu_q = 7.56$. In these comparatively thick films, the helicoids remain 
thermodynamically stable at low fields and transform
into the cone by a first-order process at the critical line $h_{\mathrm{hc}} (\nu)$.
The cone magnetization along the applied field increases linearly for increasing magnetic 
field up to the saturation at critical field $H = H_D$.

(II) $ 4.47 = \nu_p < \nu < \nu_q = 7.56$. In this case, the magnetic-field-driven evolution 
of the cone is interrupted by the first-order transition in the skyrmion lattice 
at $h_{\mathrm{hc}} (\nu) < h_q$ and the re-entrant transition 
at $h_{\mathrm{hc}} (\nu) > h_q$.
 
(III)  $ 1 < \nu < \nu_p = 4.47$. In this thickness range,  the stability area of skyrmion
lattices is separated from the low field helicoid and high field cone phases by the first
order transition lines.

\begin{figure*}
\includegraphics[width=2.0\columnwidth]{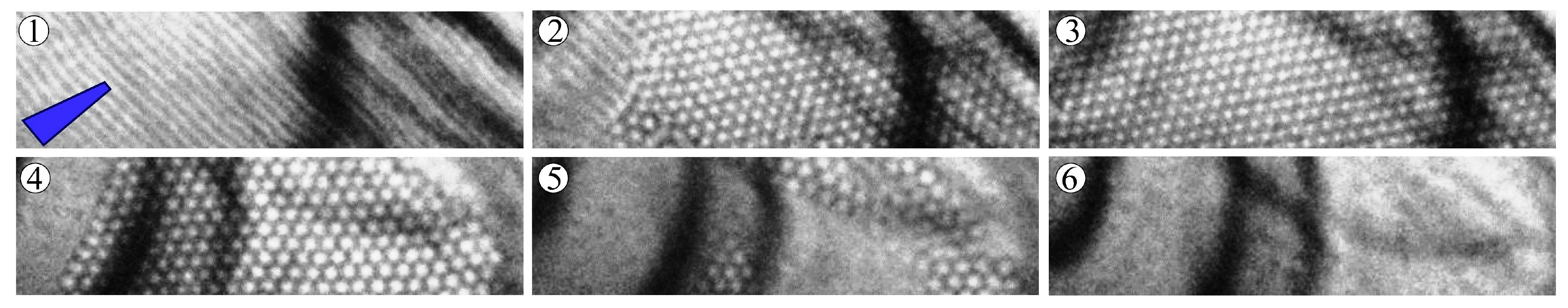}
\caption{ LTEM images of modulated phases in a FeGe wedge 
at T = 250 K and different values of the applied field
H (Oe): 130 (1), 873 (2), 1073 (3), 2215 (4), 2355 (5),
3728 (6);
fig. (3) indicates the coexisting helicoid and skyrmion lattice 
states and figs. (4), (5) the skyrmion lattice and
cone domains during the first-order phase transitions.
The image size is 3000 nm $\times$ 800 nm, the thickness varies
from 140 nm (left) to 60 nm (right).
Blue tetragons  indicate the direction of the thickness gradients
(in this and the next figures).
\label{fig:wedge1}
}
\end{figure*}

\begin{figure}
\includegraphics[width=1.0\columnwidth]{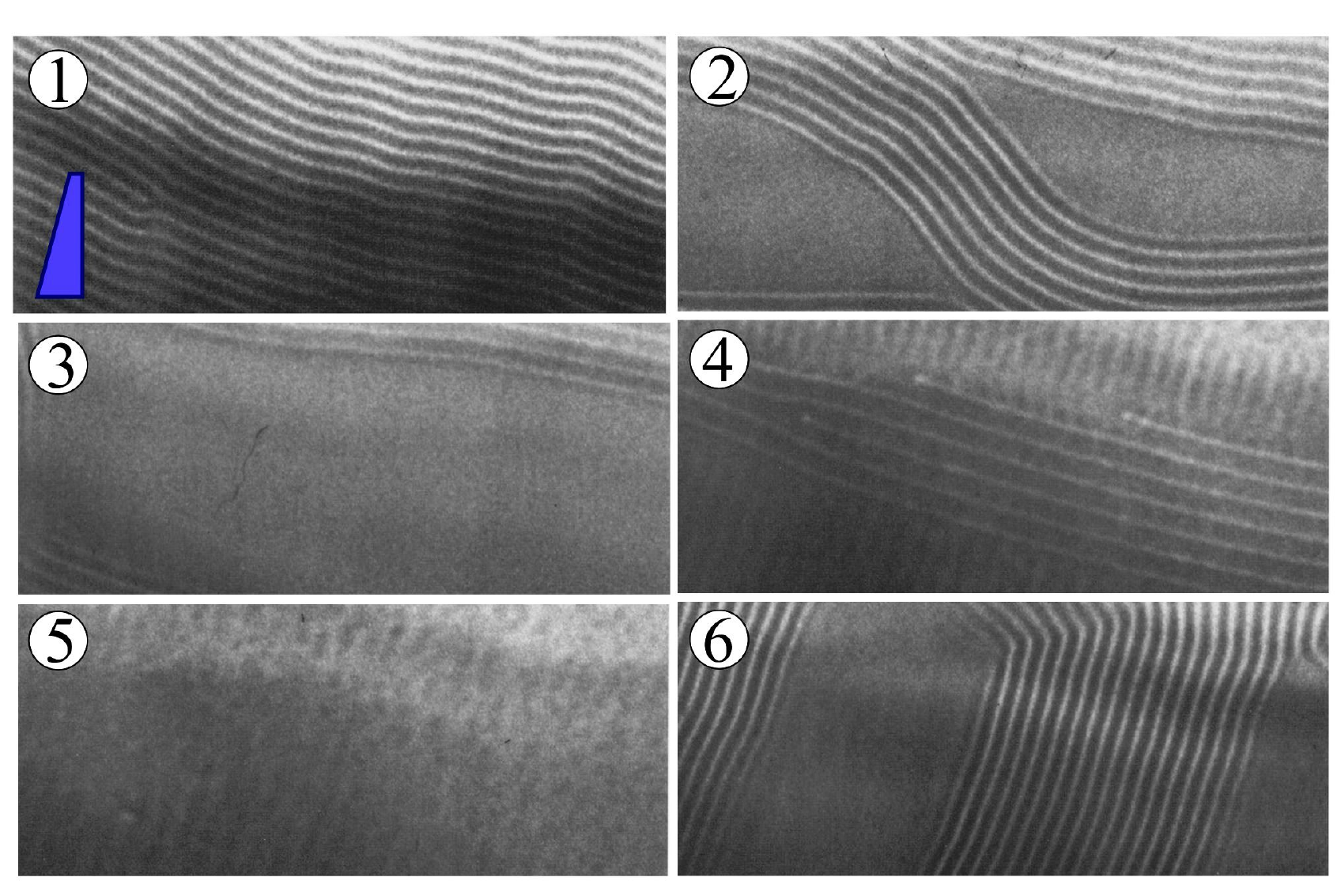}
\caption{
(color online). LTEM images of a FeGe wedge at $T = $110 K for  applied magnetic fields: 
H (Oe) =  200 (1), 1074 (2), 1460 (3), 3200 (4), 3670 (5), 200 (6) Figs. 2, 3 
show the coexisting domains of the helicoid and cone state during 
the first-order transition between these phases.
Multidomain states are restored after  descreasing 
of the applied field (fig. (6)).
The image size is 3000 nm $\times$ 1250 nm, the thickness varies
from 90 nm (bottom) to 60 nm (top).
\label{fig:wedge2}
}
\end{figure}

\subsection{ Analytical solutions for surface twists}

In cubic helimagnet films with $L \geq L_D$, 
twisted modulations in helicoids and skyrmions
($\xi (z)$)  exist only in narrow regions near 
the film surfaces $ \delta \ll L$.
This allows us to write solutions for helicoids
as $\theta = \theta (x), \psi = \pi/2 +\xi (z)$
(the $x$ axis is directed along the propagation direction),
where $\theta (x)$ is the solution homogeneous along the
$z$ axis investigated in \cite{Dz64}.
We write the solutions for a skyrmion lattice core as
 $\theta = \theta (\rho), \psi = \pi/2 +\varphi + \xi (z)$
where and  $\theta (\rho)$ is the solutions for 
skyrmions homogeneous along their axes \cite{JMMM94}. 
The energy density of the surface twists in the helicoid
(skyrmion lattice) can be reduced to the following form:
$ e_{h(s)} (\xi) = \Delta \bar{w}_{h(s)} (\xi) =
\left\langle m_x^2\right\rangle_{h(s)} \mathcal{F}_{h(s)} (\xi)$ where

\begin{eqnarray}
\textstyle
\mathcal{F}_{h(s)}(\xi)=\frac{2}{L} \int_0^{\infty}\! dz\!
 \left[ A \xi_z^2 -\!D\xi_z - K_0 \: \upsilon_{h(s)}  \sin^2 \frac{\xi}{2} \right], \:
\label{twist1}
\end{eqnarray}
$K_0$ is the effective anisotropy (\ref{cone}), and
\begin{eqnarray}
\textstyle
\left\langle m_x^2\right\rangle_h = \frac{1}{l} \int_0^l \sin^2 \theta dx, \quad
\left\langle m_x^2 \right\rangle_s = \frac{1}{\pi R_s^2} \int_0^{R_s} \sin^2 
\theta \rho d\rho, \quad
\nonumber \\
\upsilon_h = (4 L_D/l) \left\langle m_x^2\right\rangle_h^{-1}, \quad
\upsilon_s =  2 \eta_D (L_D/\pi) \left\langle m_x^2\right\rangle_s^{-1}, \quad \quad 
\nonumber 
\label{twist2}
\end{eqnarray}
$\eta_D = \frac{1}{\pi R_s^2} \int_0^{R_s} \left( \theta_{\rho} 
+\frac{1}{\rho} \sin \theta \cos \theta \right) \rho d \rho $.

The energy functional   $\mathcal{F}_{h(s)} (\xi)$  (\ref{twist1}) 
describes surface twists $\xi (z)$ in helicoids (skyrmion lattices)
and has the same  functional form as the energy functional 
for surface twists in a saturated helimagnet \cite{Meynell14}.
The Euler equation for (\ref{twist1}) can be readily solved analytically.
The equilibrium amplitude of twist  modulations $\xi (z) $ reaches 
the largest value on the film surface, 
\begin{eqnarray}
\xi_{h(s)}^{(0)}   = 2 \arcsin  \left( \upsilon_{h(s)}^{-1/2} \right)
\label{xi0}
\end{eqnarray}

and decays exponentially into the layer depth,

\begin{eqnarray}
\textstyle
\tan (\xi_{h(s)}/4 )\! =\!\tan (\xi_{h(s)}^{(0)}/4 )\:
e^{\left[-\pi \sqrt{\upsilon_{h(s)}} \left(z/L_D \right) \right]}. \ 
\label{xi}
\end{eqnarray}

Inserting (\ref{xi}) into the energy density (\ref{twist1}) leads to the following
expression for the \textit{negative} energy density contribution imposed 
by  surface twist modulations:

\begin{eqnarray}
 \textstyle
\bar{e}_{h(s)} = \underbrace{\left\langle m_x^2\right\rangle_{h(s)} 
\left[2\tan \left( \xi_{h(s)}^{(0)} /4 \right)\!-\!\xi_{h(s)}^{(0)} \right] }_{\sigma_{h(s)}}/L < 0. \; 
\label{deltaw}
\end{eqnarray}

The fractions of the \textit{negative}  surface contribution  
$\bar{e}_{h(s)} \propto 1/L$ (\ref{deltaw})
in the total energy balance increase with decreasing film thickness,
extending the stability areas of the helicoids and skyrmion lattices 
(\ref{fig:phasediagram}).

The magnetic phase diagram in Fig. \ref{fig:phasediagram} has been
derived by minimization of the simplified (``isotropic'') 
energy functional (\ref{density}) with \textit{free} 
boundary conditions. This demonstrates how a \textit{pure} geometrical factor
(confinement) influences the energetics of cubic helimagnet nanolayers 
by imposing transerverse chiral modulations (\textit{twists}) in skyrmion 
lattices and helicoids.
For the practically important thickness range $L \geq L_D$,
the influence of chiral twists can be described by the surface energy
term (\ref{deltaw}) in the cubic helimagnet energy.

\begin{figure}
\includegraphics[width=0.8\columnwidth]{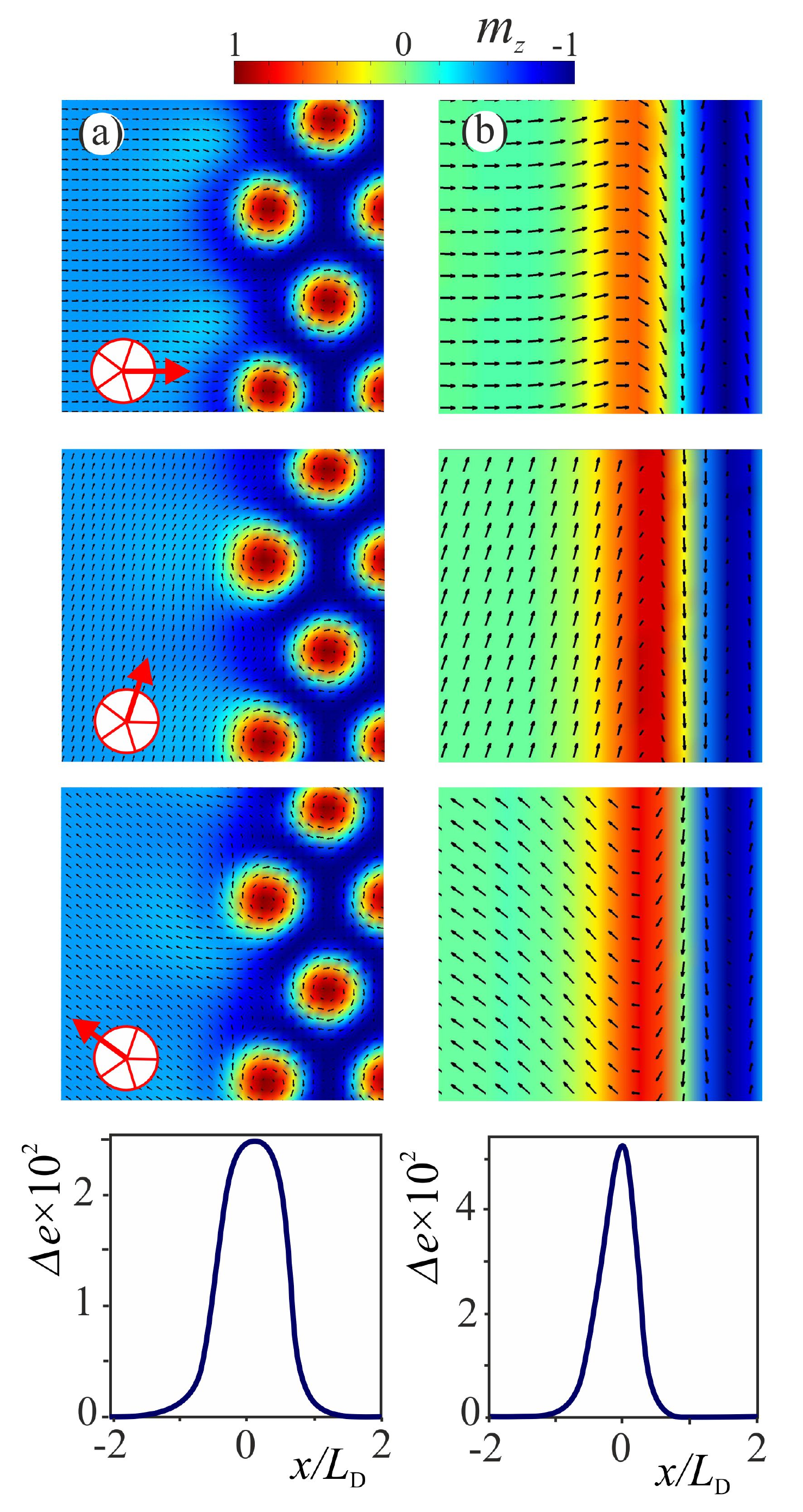}
\caption{
(color online). Contour plots show the equilibrium structures 
of domain boundaries  between the cone and the competing skyrmion lattice (a)
and helicoid phases (b) during the first-order transition. The applied
magnetic field is directed perpendicular to the figures planes.
The domain wall between the cone and skyrmion lattice is calculated at the
critical line $h_{\mathrm{sc}} (\nu)$ in the vicinity of the completion point 
$q$ ($ h = 0.40$) (a), and the domain wall between the cone and 
helicoid at the transition line $h_{\mathrm{hc}} (\nu)$ for $h = 0.01$ (b).
The bottom panel shows the reduced energy density profiles
through the domain wall thickness 
$ \Delta e (x) = |(w (x) - w_c (h))/w_c (h)|$ where $w_c (h)$ is
defined by Eq. (\ref{cone}).
\label{fig:dw2}
}
\end{figure}%

\section{Evolution of skyrmion and helical states in a FeGe wedge}

Iron monogermanide FeGe belongs to a group of non-centrosymmetric
cubic helimagnets with space group P2$_1$3 (B20-type structure) 
\cite{Lebech89,Wilhelm12}. Below the Curie temperature $T_C =278.2$ K,
FeGe is ordered into  homochiral helices with period 
$L_D = 70$ nm propagating along equivalent $<100>$ directions 
\cite{Lebech89}. Below $T_1$ = 211 K, helices propagate along
$<111>$ directions. For increasing temperature, 
the propagation directions $<100>$
are restored at $T_2$ = 245 K \cite{Lebech89}.

In bulk cubic helimagnets, one-dimensional single-harmonic 
chiral modulations (helices and cones) 
are observed as stable states over practically the entire 
region below the saturation field \cite{Lebech89}.
Contrary to bulk specimens, in free standing nanolayers of
cubic helimagnets with thickness $L \leq 120$ nm investigated by
LTEM methods, skyrmion lattices and helicoids are observed
in broad ranges of applied magnetic fields and temperatures,
while the cone phase is partially or completely suppressed 
\cite{Yu10,Yu11,Yu15}. Recent LTEM investigations represent at
extensive study of the evolution of skyrmion states
in confined cubic helimagnets (see e.g. \cite{Yu10,Yu11,Onose12,Yu15} 
and bibliography in \cite{Nagaosa14}).

In our paper we use LTEM to explore first-order phase transitions 
into the cone phase  and other specific magnetization 
processes imposed by the \textit{chiral surface twists }
(Fig. \ref{fig:phasediagram}).
 For our studies, we have prepared wedge-shape single crystal FeGe(110) films.  
FeGe single crystals were grown by a chemical vapor transport method.
A thin film specimen was made for TEM observations by using 
a focused ion beam technique. 
A series of Lorentz micrographs were taken by means of a Fresnel mode of
Lorentz microscopy with a typical defocus value of 10 micrometer
at $T = $ 110 K and 250 K in a broad range of magnetic fields applied
perpendicular to the film surface 
(Figs. \ref{fig:wedge1}, \ref{fig:wedge2}).
They clearly expose the magnetic-field-driven first-order transitions between
the basic modulated states accompanied by the formation of the multidomain patterns 
composed of domains of the competing phases.

It should be noted that at LTEM images the domains of the cone phase appear as dark
areas and cannot be distinguished from domains of the saturated state. However, in 
Figs. \ref{fig:wedge1}, \ref{fig:wedge2}, the dark domains arise at applied fields 
lower than the saturated fields (for FeGe,  the saturation field  $\mu_0 H_D = $ 0.359 T 
(\ref{period}) \cite{Lebech89,Wilhelm12}).
Moreover, according to the theoretical results \cite{Dz64,JMMM94} and experimental 
observations \cite{Togawa12,Yu10}, the magnetic-field-driven transitions of  
the helicoid and skyrmion lattice into the saturated state advance gradually 
by the extension of the modulation period and formation of isolated helicoidal kinks 
and skyrmions. 
These processes exclude the formation of multidomain patterns of the competing 
phases characteristic of the first-order transitions \cite{Hubert98}.

In Fig. \ref{fig:wedge1}, the layer thickness varies from $L = 140$ nm ($\nu =2$)
at the left edge to $L = 60$ nm ($\nu =0.86$) at the right edge.
In the calculated phase diagram, this thickness interval  ($0.86 < \nu < 2$)
belongs to \textit{area III} characterized  by the first-order transitions between
the helicoid and skyrmion lattice at the lower field, $h_{hs} (\nu)$, and between
the skyrmion lattice and cone at higher field, $h_{sc} (\nu)$
(Fig. \ref{fig:phasediagram}). Both these phase transitions are clearly 
observed in  Fig. \ref{fig:wedge1}.
Because the transition field $h_{sc} (\nu)$ has lower values for larger $\nu$, 
initially the cone phase nucleates at the thicker edge of the film 
(Figs. \ref{fig:wedge1}, (4)) and expands to the thinner part with 
an increasing applied field (Figs. \ref{fig:wedge1}, (5)).

The LTEM images derived at $T = $ 	110 T (Fig. \ref{fig:wedge2}) 
have been done for a wedge area belonging to the same thickness interval 
as that in Fig. \ref{fig:wedge1} with the thickness variation from 
$L=$ 90 nm ($\nu =1.29$) at the bottom edge to $L = 60$ nm ($\nu =0.86$) 
at the top edge. However, the magnetization evolution
differs drastically from that observed at higher temperature. 
In this case, a skyrmion lattice does not arise, instead the helicoid 
directly transforms into the cone phase at a considerably lower field 
by a first-order process (Fig. \ref{fig:wedge2} (2), (3) ).
In the $(\nu, h)$ phase diagram  (Fig. \ref{fig:phasediagram}) 
such a magnetization evolution occurs in the \textit{area I}  
for $\nu > \nu_q = 7.56$. 
The suppression of skyrmion lattices and helicoids at lower temperatures 
is characteristic for free-standing cubic helimagnet nanolayers 
\cite{Yu11,Yu15}. 
Particularly, at $T$ = 110 K, the skyrmion lattices arise in FeGe free-standing 
layers only when their thickness is smaller than 35 nm \cite{Yu11}.
This effect can be understood if we assume that the surface energy imposed 
by chiral twists $\sigma_{h(s)}$ (\ref{deltaw}) decreases with decreasing 
temperatures. 
As a result, at lower temperatures the existence area of skyrmion lattices 
in the ($\nu, h$) phase diagram (\ref{fig:phasediagram}) would be shifted 
into the region of lower $\nu$.

Finally we consider specific domain wall separating domains of 
the competing modulated phases during the first-order transition. 
The transition between the helicoid and skyrmion lattice occurs
in bulk and confined chiral helimagnets \cite{JMMM94,Rybakov13}. 
Domain walls between the coexisting helicoids and skyrmion lattices 
\cite{JMMM94} have been observed in confined cubic helimagnets 
\cite{Yu10,Rajeswari15,Nagao15} and FePd/Ir bilayers \cite{Romming13}.

In the multidomain patterns in Figs. \ref{fig:wedge1}, \ref{fig:wedge2} 
the domain boundaries between the cone phase and the helicoids
and skyrmion lattices represent specific transitional areas providing
the compatibility of the chiral modulations along the applied field 
(the cone) with the in-plane modulated phases. The contour plots
in Fig. \ref{fig:dw2} describe the equilibrium structure of isolated
domain walls between the cone and skyrmion lattice calculated
for  $h = 0.4$ and between the cone and helicoid at $h=0$.
These calculations have been carried out for homogeneous along
the film thickness domains of the helicoid and skyrmion lattice 
phases.
The energy density profiles of the domain walls in Fig. \ref{fig:dw2}
show that the potential barriers separated the equilibrium modulated
phases in domain are estimated as 
$\Delta w_{max} (h) = \Delta e (0) |w_c(h)| \approx 10^{-2} K_0$.

\section{Conclusions}

The results of micromagnetic calculations 
for confined chiral modulations and LTEM investigations
of magnetic states in a FeGe wedge demonstrate that
chiral surface twists provide the stabilization mechanism 
for helicoids and skyrmion lattices in free standing
cubic helimagnet films.
For a practically important thickness range
$L \geq L_D$, chiral twist modulations have sizable values only near the
film surfaces and can be described analytically as localized surface states
exponentially decaying into the film depth (Eqs. (\ref{xi0}), (\ref{xi})).
The stabilization energy in this case is described by the surface energy
contributions (\ref{deltaw}).

The solutions minimizing the energy functional (\ref{density}) 
with free boundary conditions describe chiral modulations 
imposed solely by the \textit{geometrical confinement} 
and expose three basic types of the magnetization processes 
in cubic helimagnet nanolayers (Fig. \ref{fig:phasediagram}). 
In real system, however, the confined chiral modulations arise
as a result of the interplay between the stabilization mechanism 
imposed by the geometrical confinement and other physical factors, 
such as intrinsic cubic anisotropy and induced volume and surface 
uniaxial anisotropy, internal and surface demagnetization 
effects.
Our findings provide a conceptional basis for detailed experimental and
theoretical investigations of the complex physical processes underlying the
formation of skyrmion lattices and helicoids in confined noncentrosymmetric
magnets.


\begin{thebibliography} {99}

\bibitem{Dz64}  
I.\ E.\ Dzyaloshinskii, Sov.\ Phys.\ JETP {\textbf{19}}, 960 (1964), {\textbf{20}}, 665 (1964).

\bibitem{JETP89}
A. N. Bogdanov and D. A. Yablonskii, Sov. Phys. JETP \textbf{68}, 101 (1989), \
 \blue{http://www.jetp.ac.ru/cgi-bin/e/index/r/95/1/p178?a=list}.

\bibitem{JMMM94}
A. Bogdanov and A. Hubert, 
J. Magn. Magn. Mater. {\textbf{138}}, 255 (1994), 
{\textbf{195}}, 182 (1999).

\bibitem{Bak80}
P. Bak and M. H. Jensen, J.Phys. C {\textbf{13}}, L881 (1980).

\bibitem{Butenko10}
A. B. Butenko, A. A. Leonov, U. K. R\"{o}{\ss}ler, A. N. Bogdanov,
 Phys. Rev. B {\textbf{82}}, 052403 (2010).
\bibitem{Wilson14}
M. N. Wilson, A. B. Butenko, A. N. Bogdanov, and T. L. Monchesky 
Phys. Rev. B {\textbf{89}}, 094411 (2014).



\bibitem{Yu10}
X. Z. Yu, Y. Onose, N. Kanazawa, J. H. Park, J. H. Han,
Y. Matsui, N. Nagaosa, and Y. Tokura,  
Nature {\textbf{465}}, 901 (2010).

\bibitem{Yu11}
X. Z. Yu, N. Kanazawa, Y. Onose, K. Kimoto, W. Z. Zhang, S.
Ishiwata, Y. Matsui, and Y. Tokura, 
Nature Mat. {\textbf{10}}, 106 (2011).

 
\bibitem{Yu15}
X. Z. Yu, A. Kikkawa, D. Morikawa, K. Shibata, Y. Tokunaga,
Y. Taguchi, and Y. Tokura,  
Phys. Rev. B {\textbf{91}}, 054411 (2015).

\bibitem{Wilson12} 
M. N. Wilson, E. A. Karhu, A. S. Quigley, U. K. R\"{o}{\ss}ler, A. B. Butenko, 
A. N. Bogdanov, M. D. Robertson, and T. L. Monchesky,  
Phys. Rev. B \textbf{86}, 144420 (2012).

\bibitem{Yokouchi15} 
T. Yokouchi, N. Kanazawa, A. Tsukazaki, Y. Kozuka, A. Kikkawa, Y. Taguchi, 
M. Kawasaki, M. Ichikawa, F. Kagawa, Y. Tokura,
J. Phys. Soc. Jpn. \textbf{84}, 104708 (2015).

\bibitem{Huang12} 
S. X. Huang and  C. L. Chien, 
Phys. Rev. Lett. {\textbf{108}}, 267201 (2012).

\bibitem{Karhu10}
E. Karhu, S. Kahwaji, T. L. Monchesky, C. Parsons, M. D.
Robertson, and C. Maunders, {\textbf{82}}, 184417 (2010);
 E. A. Karhu, S. Kahwaji, M. D. Robertson, H. Fritzsche, B. J.
Kirby, C. F. Majkrzak, and T. L. Monchesky, 
Phys. Rev. B {\textbf{84}}, 060404 (2011).
   

\bibitem{Togawa12}
Y. Togawa, T. Koyama, K. Takayanagi, S. Mori, Y. Kousaka,J. Akimitsu, 
S. Nishihara, K. Inoue, A. S. Ovchinnikov, and J. I. Kishine, 
Phys. Rev. Lett. {\textbf{108}}, 107202 (2012).

\bibitem{Hubert98}
A. Hubert, R. Sch\"{a}fer, \textit{Magnetic Domains} (Springer, Berlin, 1998);
V. G. Baryakhtar, A. N. Bogdanov, D. A. Yablonskii,
 Sov. Phys. Usp. {\textbf{31}}, 810 (1988).



\bibitem{Wilson13} 
M. N. Wilson, E. A. Karhu, D. P. Lake, A. S. Quigley, 
A. N. Bogdanov, U. K. R\"{o}{\ss}ler, T. L. Monchesky, 
Phys. Rev. B  {\textbf{88}}, 214420 (2013).

\bibitem{Rybakov13}
F. N. Rybakov, A. B. Borisov, A. N. Bogdanov,
Phys. Rev. B {\textbf{87}}, 094424 (2013);
F. N. Rybakov, A. B. Borisov, S. Bl\"{u}gel, N. S. Kiselev,
Phys. Rev. Lett {\textbf{115}}, 117201 (2015).

\bibitem{Meynell14}
S. A. Meynell, M. N. Wilson, H. Fritzsche, A. N. Bogdanov, and T. L. Monchesky, 
Phys. Rev. B {\textbf{90}}, 014406 (2014).

\bibitem{Neel54}
L. N$\acute{e}$el, J. Phys. Rad.  {\textbf{15}}, 225 (1954).

\bibitem{Lebech89}
B. Lebech, J. Bernhard, and T. Freltoft,
J. Phys.: Condens. Matter {\textbf{1}}, 6105 (1989).

\bibitem{Wilhelm12}
H.Wilhelm, M. Baenitz, M. Schmidt, C. Naylor, R. Lortz, 
U. K. R\"{o}{\ss}ler, A. A. Leonov, and A. N. Bogdanov,
J. Phys.: Condens. Matter  {\textbf{24}}, 294204 (2012).

\bibitem{Nagaosa14}
N. Nagaosa and Y. Tokura, Nat. Nanotech.  {\textbf{8}}, 899 (2013).

\bibitem{Rajeswari15}
J. Rajeswari,H. Ping, G. F. Mancini, Y. Murooka, T. Latychevskaia, 
D. McGrouther, M. Cantoni, E. Baldini, J. S. White, A. Magrez, 
T. Giamarchi, H. M. Roennow, F. Carbone,
PNAS (2015). 

 \bibitem{Nagao15}
M. Nagao, Y. So, H. Yoshida, K. Yamaura, T. Nagai,
T. Hara, A. Yamazaki, and K. Kimoto,
  Phys. Rev. B {\textbf{92}}, 140415 (2015).

\bibitem{Onose12}
Y. Onose, Y. Okamura, S. Seki, S. Ishiwata, and Y. Tokura,
Phys. Rev. Lett. {\textbf{109}}, 037603 (2012).

\bibitem{Romming13}
N. Romming, C. Hanneken, M. Menzel, J. E. Bickel, B.
Wolter, K. von Bergmann, A. Kubetzka, and R. Wiesendanger,
Science {\textbf{341}}, 636 (2013).





\end{thebibliography}
\end{document}